\documentclass[
aps,
preprint,
superscriptaddress,
onecolumn,
longbibliography]{revtex4-1}
\usepackage{graphicx}
\usepackage{latexsym}
\usepackage{amssymb}
\usepackage{amsmath}
\usepackage{amsfonts}
\usepackage{upgreek}
\usepackage{bm}
\usepackage{multirow}
\usepackage{enumitem}
\usepackage{color}
\usepackage{hyperref}
\usepackage{lipsum}
\usepackage{tabularx}
\begin{document}
\title{Active learning of reactive Bayesian force fields: Application to heterogeneous hydrogen-platinum catalysis dynamics}

\author{Jonathan Vandermause}
\affiliation{Department of Physics, Harvard University, Cambridge, Massachusetts 02138, USA}
\affiliation{John A. Paulson School of Engineering and Applied
Sciences, Harvard University, Cambridge, MA 02138, USA}

\author{Yu Xie}
\affiliation{John A. Paulson School of Engineering and Applied
Sciences, Harvard University, Cambridge, MA 02138, USA}

\author{Jin Soo Lim}
\affiliation{Department of Chemistry and Chemical Biology, Harvard University, Cambridge, MA 02138, USA}

\author{Cameron J. Owen}
\affiliation{Department of Chemistry and Chemical Biology, Harvard University, Cambridge, MA 02138, USA}

\author{Boris Kozinsky}
\affiliation{John A. Paulson School of Engineering and Applied
Sciences, Harvard University, Cambridge, MA 02138, USA}
\affiliation{Bosch Research, Cambridge, MA 02139, USA}

\date{\today}

\begin{abstract}
Accurate modeling of chemically reactive systems has traditionally relied on either expensive \textit{ab initio} approaches or flexible bond-order force fields such as ReaxFF that require considerable time, effort, and expertise to parameterize. Here, we introduce FLARE++, a Bayesian active learning method for training reactive many-body force fields “on the fly” during molecular dynamics (MD) simulations. During the automated training loop, the predictive uncertainties of a sparse Gaussian process (SGP) force field are evaluated at each timestep of an MD simulation to determine whether additional \textit{ab initio} data are needed. Once trained, the SGP is mapped onto an equivalent and much faster model that is polynomial in the local environment descriptors and whose prediction cost is independent of the training set size. We apply our method to a canonical reactive system in the field of heterogeneous catalysis, hydrogen splitting and recombination on a platinum (111) surface, obtaining a trained model within three days of wall time that is twice as fast as a recent Pt/H ReaxFF force field and considerably more accurate. Our method is fully open source and is expected to reduce the time and effort required to train fast and accurate reactive force fields for complex systems.
\end{abstract}
\maketitle

\section{Introduction}
Accurate modeling of chemical reactions is a central challenge in computational physics, chemistry, and biology, and is at the heart of computer-aided design of covalent drugs \cite{singh2011resurgence} and next-generation catalysts for CO$_2$ reduction \cite{friend2017heterogeneous}. Reactive molecular dynamics (MD) is an essential tool in addressing these challenging problems \cite{van2001reaxff}. By directly simulating the motion of individual atoms without fixing any chemical bonds, reactive MD makes it possible to uncover subtle reaction mechanisms with atomic-scale resolution and to predict reaction rates that can complement experimental studies \cite{mihalovits2020affinity}.

Reactive MD requires a flexible model of the potential energy surface (PES) of the system that is both (i) sufficiently \textit{accurate} to faithfully describe the making and breaking of chemical bonds, and also (ii) sufficiently \textit{cheap} to open up the long timescales often associated with rare reactive events. Although for small systems it is possible to evaluate the PES at each timestep of the simulation using accurate \textit{ab initio} methods such as density functional theory (DFT) \cite{kohn1965self} or post-Hartree-Fock \cite{bartlett2007coupled}, these methods scale nonlinearly with the number of electrons and are therefore difficult to extend to systems containing more than a few hundred atoms or to timescales exceeding a few hundred picoseconds.

For many nonreactive systems, a viable alternative to \textit{ab initio} MD is to parameterize a force field for the system of interest, which typically scales linearly with the number of atoms and is often orders of magnitude cheaper than \textit{ab initio} models of the PES. However, many traditional force fields, such as the AMBER and CHARMM models extensively used in biomolecular simulations \cite{wang2004development, vanommeslaeghe2010charmm}, explicitly fix the chemical bonds in the system and are therefore not suitable for describing chemical reactions. While the more flexible ReaxFF force field is capable of describing bond breaking and formation and has been applied to a wide range of reactive systems over the last two decades \cite{van2001reaxff, senftle2016reaxff}, ReaxFF models can significantly deviate from the \textit{ab initio} PES \cite{bartok2018machine} and usually require expert hand-tuning for each system.

Machine-learned (ML) force fields have emerged in the last decade as a powerful tool for building linear-scaling models of the PES that achieve near-DFT accuracy \cite{behler2007generalized, bartok2010gaussian, thompson2015spectral, shapeev2016moment} and have recently been applied to reactive systems \cite{zeng2020complex, yang2021using, young2021transferable}. The most common approach to fitting these models involves manually building a database of \textit{ab initio} calculations targeting the system of interest. Potential energies, atomic forces, and virial stress tensors from these calculations are then used to train a flexible regression model that maps descriptors of atomic environments onto local energies, which are summed to predict the total potential energy of the system. This manual approach to training ML force fields has been widely successful in describing a range of bulk materials \cite{szlachta2014accuracy, deringer2017machine, behler2017first, deringer2018data, bartok2018machine, zhang2018end} and organic molecules \cite{schutt2018schnet, christensen2020fchl, chmiela2019sgdml}, and has in some cases revealed new insights into atomistic phenomena \cite{lim2020evolution, cheng2020evidence, deringer2021origins}. However, manual training can require months of effort and considerable expertise and computing resources, and is particularly challenging to apply to reactive systems, where it is often not known in advance what the relevant transition paths are and how extensively they need to be sampled in the training database.

A powerful alternative to training ML force fields manually is to generate the training set autonomously using active learning \cite{artrith2012high, rupp2014machine, smith2018less, zhang2019active, bernstein2019novo, sivaraman2020machine}. In this approach, model errors or uncertainties are used to decide if a candidate structure should be added to the training set, in which case an \textit{ab initio} calculation is performed and the model is updated. A particularly promising active learning approach that enables rapid force field development for target systems involves training the model ``on the fly'' during molecular dynamics, in which the model proposes MD steps and is updated only when model uncertainties exceed a chosen threshold \cite{li2015molecular, podryabinkin2017active, jinnouchi2020fly}. On-the-fly learning has been successfully applied in the last year to a range of nonreactive systems and phenomena, including phase transitions in hybrid perovskites \cite{jinnouchi2019phase}, melting points of solids \cite{jinnouchi2019fly}, superionic behavior in silver iodide \cite{vandermause2020fly}, the 2D-to-3D phase transition of stanene \cite{xie2021bayesian}, and lithium-ion diffusion in solid electrolytes \cite{hajibabaei2020towards}, and has recently been extended by Young \textit{et al.}\ to chemical reactions within the Gaussian approximation potential (GAP) framework \cite{young2021transferable}. The kernel-based GAP force field is highly accurate, but has a prediction cost that grows linearly with the size of the sparse set and is several orders of magnitude more expensive than traditional force fields such as ReaxFF \cite{plimpton2012computational}.

Here, we develop a method for training reactive many-body force fields on the fly and accelerating the resulting models so that their prediction speed exceeds that of ReaxFF. We use Bayesian uncertainties of a GAP-style sparse Gaussian process (SGP) force field to decide which structures to include in the training set, and show that the prediction cost of the model can be made independent of the training set size by mapping it onto an equivalent parametric model that is polynomial in the local environment descriptors. We apply the method to a canonical reactive system in the field of heterogeneous catalysis, hydrogen gas on a platinum (111) surface \cite{christmann1976adsorption}, and use the accelerated force field to perform large-scale reactive MD of hydrogen splitting and recombination. The model is more than twice as fast as a recent ReaxFF force field developed for the same system \cite{gai2016atomistic}, and is considerably more accurate, giving an activation energy of hydrogen turnover in close agreement with experiment.

\section{Results}
\subsection{Active learning and acceleration of many-body force fields}

\begin{figure*}
	\centering
    \includegraphics{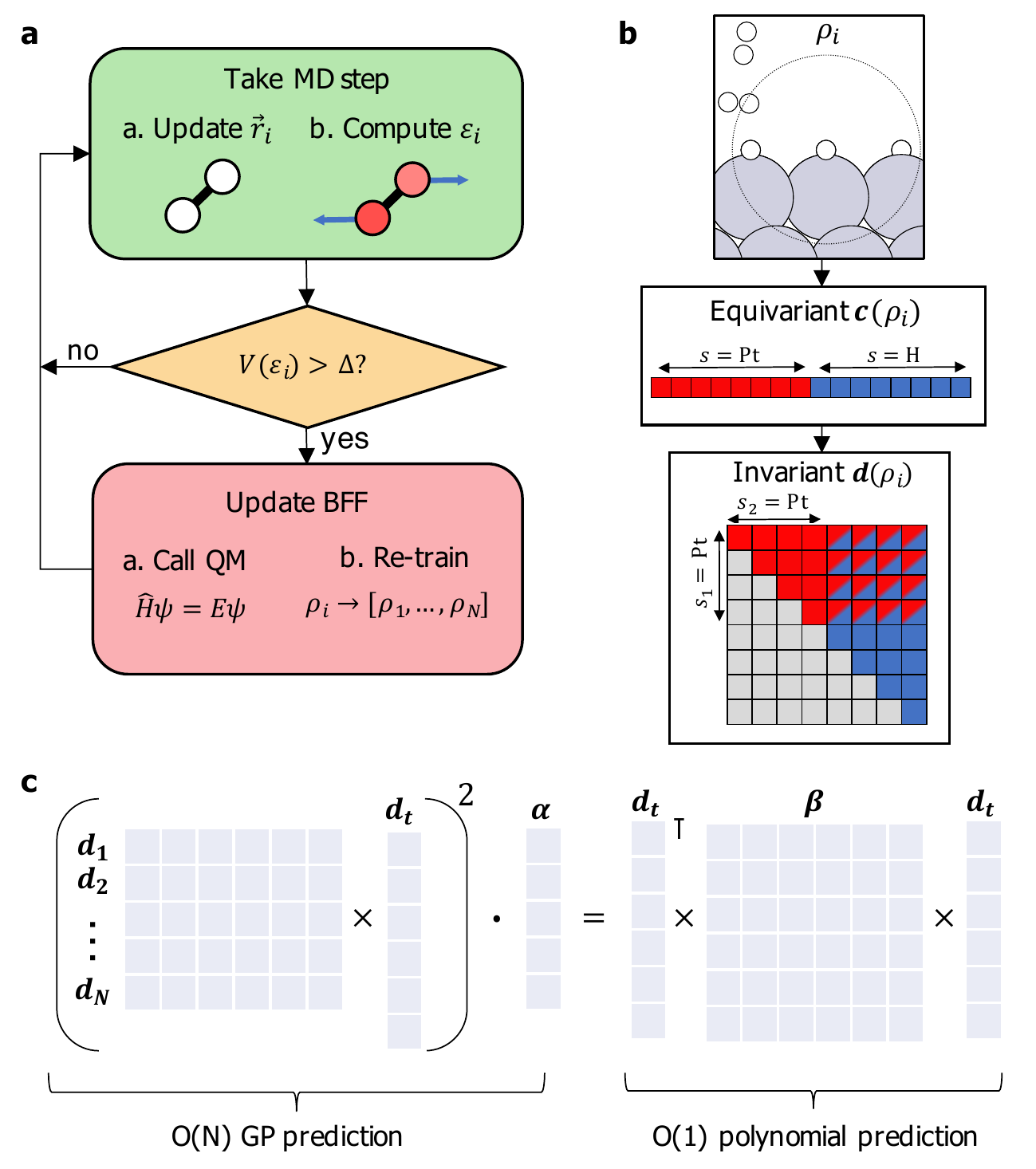}
	\caption{On-the-fly training and acceleration of many-body force fields. (a) Overview of the training procedure. At each step in the simulation, local energies, forces, and stresses are computed with the sparse GP. If the uncertainty on a local energy exceeds the chosen tolerance $\Delta_{\text{DFT}}$, DFT is called and the model is updated. (b) Mapping of local environments $\rho_i$ onto multi-element descriptors derived from the atomic cluster expansion. The environment is first mapped onto a covariant descriptor $\mathbf{c}_i$, products of which are used to compute the rotationally invariant descriptor $\mathbf{d}_i$ that serves as input to the model. (c) Mapping of a $\xi=2$ SGP force field onto an equivalent quadratic model. The cost of SGP prediction scales linearly with the number of sparse environments $N_S$, while the cost of prediction with the corresponding polynomial model is independent of $N_S$.}
	\label{fig1}
\end{figure*}

Fig.\ \ref{fig1} presents an overview of the FLARE++ method for automatically training and accelerating many-body Bayesian force fields. Our on-the-fly training loop, sketched in Fig.\ 1(a), is driven by a GAP-style SGP that provides Bayesian uncertainties on model predictions. This builds on our previous on-the-fly training method, which was restricted to two- and three-body interactions and used a more expensive exact Gaussian process \cite{vandermause2020fly}. The goal is to automatically construct both the training set and sparse set of the SGP, where the latter is a collection of representative local environments that are summed over at test time to make predictions and evaluate uncertainties (see Methods). The training loop begins with a call to DFT on an initial structure of atoms, which is used to initialize the training set and serves as the first frame of the simulation. For each MD step, the SGP predicts the potential energy, forces, and stresses of the current structure of atoms and assigns Bayesian uncertainties to local energy predictions $\varepsilon$. These uncertainties take the form of a scaled predictive variance $\widetilde{V}(\varepsilon)$ lying between $0$ and $1$ and are defined to be independent of the hyperparameters of the SGP kernel function (see Eq.\ (\ref{scaled_var}) in Methods). This is found to provide a robust measure of distance between environments $\rho_i$ observed in the training simulation and the environments $\rho_s$ stored in the sparse set of the SGP. If the uncertainty on each local energy prediction is below a chosen tolerance $\Delta_{\text{DFT}}$, the predictions of the SGP are accepted and an MD step is taken with the model. When the uncertainty of a local energy prediction exceeds the tolerance, the simulation is halted and a DFT calculation on the current structure is performed. The computed DFT energy, forces, and stresses are then added to the training database of the SGP, and local environments $\rho_i$ with uncertainty above an update tolerance $\Delta_{\text{Update}} \le \Delta_{\text{DFT}}$ are added to the sparse set. This active selection of sparse points reduces redundancy in the model by ensuring that only sufficiently novel local environments are added to the sparse set. It also helps reduce the cost of SGP predictions, which scale linearly with the size of the sparse set (see Eq.\ (\ref{en_pred}) of Methods).

The accuracy of the learned force field is in large part determined by the sophistication of the descriptor vectors assigned to local atomic environments. Our procedure for mapping local environments $\rho_i$ onto symmetry-preserving many-body descriptors $\mathbf{d}_i$ draws on the atomic cluster expansion introduced by Drautz \cite{drautz2019atomic, drautz2020atomic} and is sketched in Fig.\ 1b. A rotationally covariant descriptor $\mathbf{c}_i$, defined in Eq.\ (\ref{cov}) of Methods, is first computed by passing interatomic distance vectors through a basis set of radial functions and spherical harmonics and summing over all neighboring atoms of a particular species. Rotationally invariant contractions of the tensor product $\mathbf{c}_i \otimes \mathbf{c}_i$ are then collected in a vector $\mathbf{d}_i$ of many-body invariants, which serves as input to the SGP model. The vector $\mathbf{d}_i$ corresponds to the $B_2$ term in the multi-element atomic cluster expansion and is closely related to the SOAP descriptor. As discussed in \cite{drautz2019atomic}, its elements can be expressed as sums over three-body contributions and therefore capture angular information about neighboring atoms despite being computed with a single loop over neighbors.

Once sufficient training data have been collected, we map the trained SGP model onto an equivalent and faster polynomial model whose prediction cost is independent of the size of the sparse set, as shown schematically in Fig.\ 1c. This mapping draws on a duality in machine learning between kernel-based models on the one hand, which make comparisons with the training set at test time and are therefore ``memory-based'', and linear parametric models on the other, which are linear in the features and do not depend explicitly on the training set once the model is trained \cite{bishop2006pattern}. As shown in Methods, mean predictions of an SGP trained with a normalized dot product kernel raised to an integer power $\xi$ can be evaluated as a polynomial of order $\xi$ in the descriptor vector $\mathbf{d}_i$. The integer $\xi$ determines the body-order of the learned force field, which we define following \cite{glielmo2018efficient} to be the smallest integer $n$ such that the derivative of the local energy with respect to the coordinates of any $n$ distinct neighboring atoms vanishes. The simplest case, $\xi=1$, corresponds to a model that is linear in $\mathbf{d}_i$, and since the elements of $\mathbf{d}_i$ are sums of three-body contributions, the resulting model is formally three-body. $\xi=2$ models are quadratic in $\mathbf{d}_i$ and therefore five-body, with general $\xi$ corresponding to a $(2\xi+1)$-body model.

We evaluate the performance of different $\xi$ by comparing the log marginal likelihood $\mathcal{L}(\mathbf{y}|\xi)$ of SGP models trained on the same structures. $\mathcal{L}$ quantifies the probability of the training labels $\mathbf{y}$ given a particular choice of hyperparameters and can be used to identify hyperparameters that optimally balance model accuracy and complexity (see Eq.\ (\ref{log_like}) for the definition of $\mathcal{L}$). For the Pt/H system, we find that $\xi=2$ has a considerably higher likelihood than $\xi=1$ but nearly the same likelihood as $\xi=3$, with diminishing $\mathcal{L}(\mathbf{y}|\xi)$ for $\xi > 3$ (see Supplementary Figure 1). We therefore choose to train five-body $\xi=2$ models in practice, allowing local energies to be evaluated as a simple vector-matrix-vector product $\varepsilon(\rho_i) = \mathbf{d}_i \boldsymbol{\beta} \mathbf{d}_i$ whose gradients can be rapidly computed. We have implemented this quadratic model as a custom pair style in the LAMMPS code, and find a nearly twenty-fold acceleration over standard SGP mean prediction (Supplementary Figure 4). Remarkably, the resulting model is more than twice as fast as a recent Pt/H ReaxFF force field \cite{gai2016atomistic}, making it possible to perform ML-driven reactive MD simulations that are more efficient than their classical counterparts.

\subsection{On-the-fly training of a reactive Pt/H force field}
\begin{figure*}
	\centering
    \includegraphics{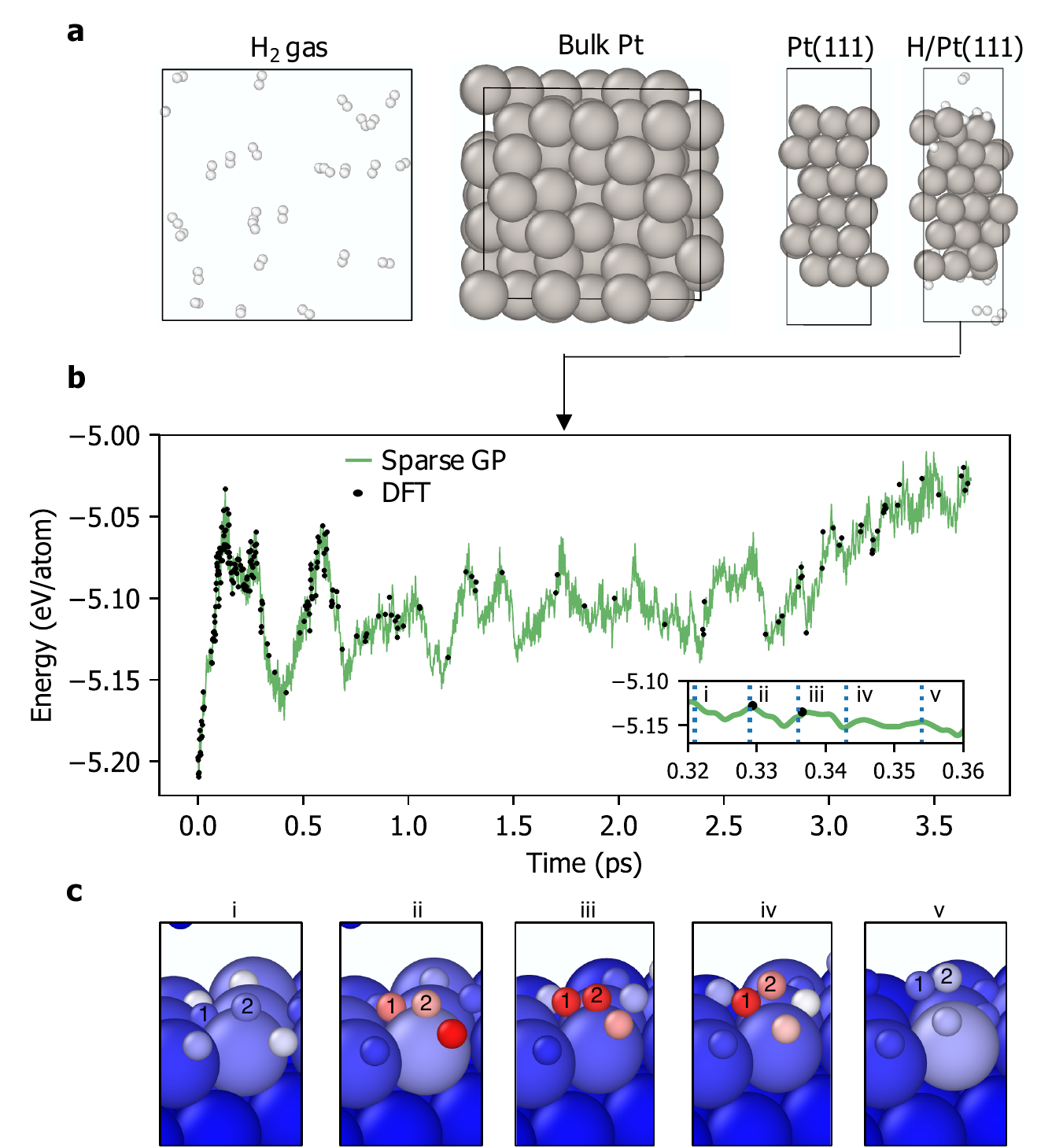}
	\caption{On-the-fly training of a reactive Pt/H force field. (a) Snapshots from the four independent on-the-fly training simulations targeting (from left to right) gas-phase hydrogen, bulk platinum, surface platinum in the (111) orientation, and hydrogen interactions with the (111) platinum surface. (b) Energy versus time during the Pt/H(111) training simulation, with SGP predictions shown in green and calls to DFT shown as black dots. The inset zooms in on model predictions during the first desorption event observed at ${t \approx 0.3 \text{ ps}}$. (c) MD snapshots of the first desorption event. Atoms are colored by the uncertainty of the local energy prediction, with red corresponding to the uncertainty tolerance $\Delta_{\text{DFT}}$.}
	\label{fig2}
\end{figure*}

Fig.\ \ref{fig2} presents our procedure for training a reactive Pt/H force field on the fly.  We performed four independent on-the-fly simulations targeting gas-phase hydrogen, surface and bulk platinum, and hydrogen adsorption and recombination on the platinum (111) surface, with example structures from each simulation shown in Fig.\ \ref{fig2}(a). Each training simulation was performed from scratch, without any data initially stored in the training set of the SGP. As reported in Table \ref{training_stats}, the single-element training simulations were each completed in less than two hours of wall time on 32 CPUs, with $24$ DFT calls made during the H$_2$ gas simulation and only $4$ and $6$ calls made during the surface and bulk platinum simulations, respectively. The majority of DFT calls were made during the reactive Pt/H training simulation, with $216$ calls made in total during the $3.7 \text{ ps}$ simulation, resulting in nearly 50,000 training labels in total (consisting of the potential energy, atomic forces, and six independent components of the stress tensor from each DFT calculation). Potential energy predictions made during the run are plotted in Fig.\ \ref{fig2}b, showing excellent agreement between the SGP and DFT to within 1 meV/atom. (Similar plots for the H$_2$ gas, bulk Pt, and Pt(111) training simulations are reported in Supplementary Figure 2.) The Pt/H training simulation was initialized with five randomly oriented H$_2$ molecules in the gas phase and with one of the two Pt surfaces fully covered with H adsorbates bound to the top site, with the temperature set to $1500$ K to facilitate rare recombination events on the Pt surface. The first recombination event in the simulation occurred at $t \approx 0.3 \text{ ps}$ and is shown as a sequence of MD snapshots in Fig.\ \ref{fig2}c. Here each atom is colored by the uncertainty of its local energy, with blue corresponding to negligible uncertainty and red corresponding to uncertainty near the DFT threshold $\Delta_{\text{DFT}}$. The formation of the chemical bond between hydrogen adsorbates 1 and 2 in Fig.\ \ref{fig2}(c) triggers two calls to DFT (corresponding to frames (ii) and (iii) in the figure), illustrating the model's ability to detect novel environments and automatically incorporate them into the training set.

\begin{figure*}
	\centering
    \includegraphics{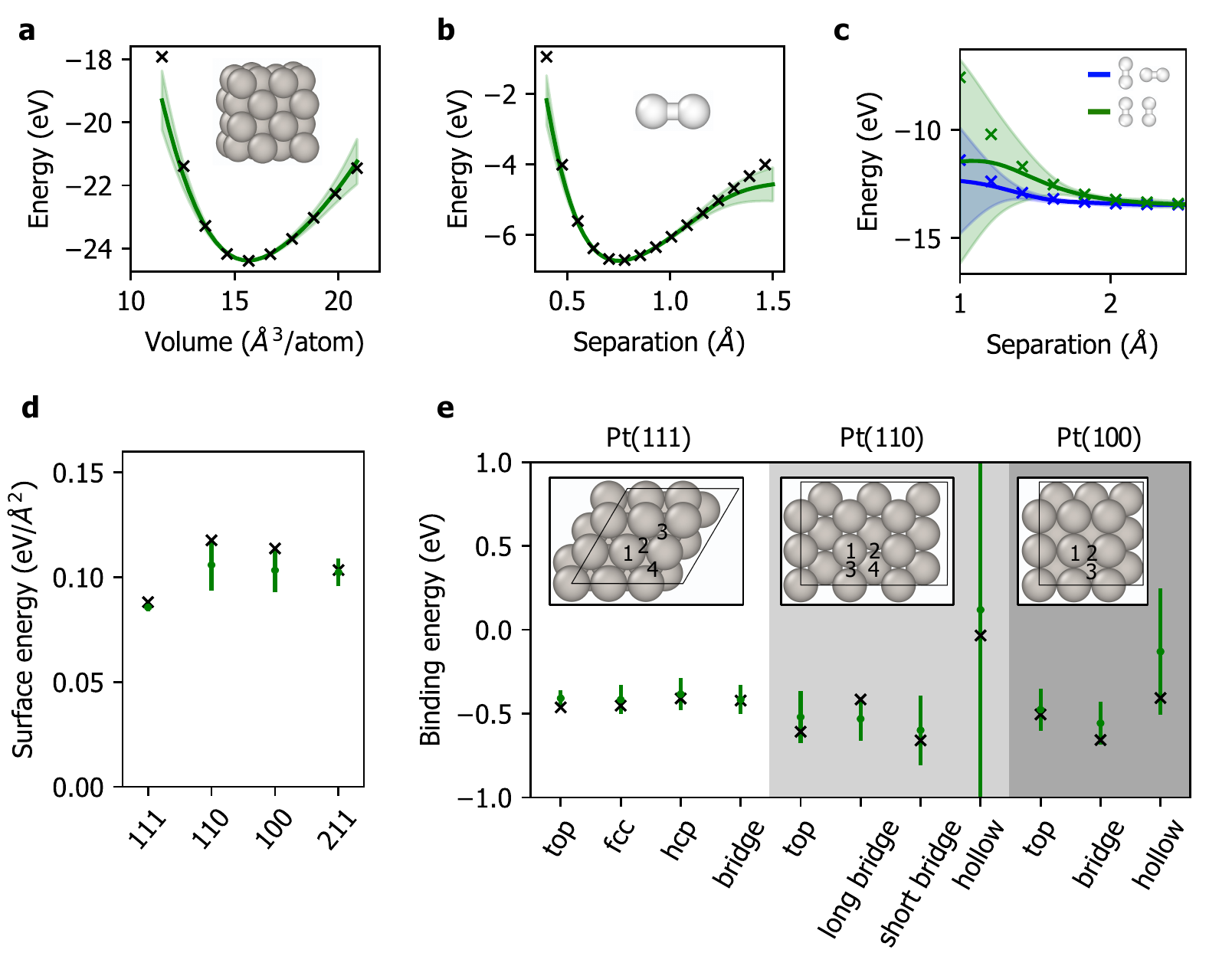}
	\caption{Validation of the trained Pt/H force field. In each plot, shaded regions indicate 99\% confidence regions of the SGP. (a) Energy versus volume of bulk Pt. (b) Energy of a single H$_2$ molecule as a function of the separation between the two H atoms. (c) Energy versus separation of two H$_2$ molecules oriented perpendicular (blue) and parallel (green) to each other. (d) Surface energies of Pt in the (111), (110), (100) and (211) orientations. (e) H binding energies for several Pt surface orientations and adsorption sites.}
	\label{fig3}
\end{figure*}

We pool together all the structures and sparse environments collected in the four independent training simulations to construct the final trained model, which we validate on a range of properties. Our goal is to obtain a model that achieves good performance not only in predicting energies, forces and stresses during finite-temperature MD simulations---quantities that the model was directly trained on---but also in predicting fundamental properties of Pt, H$_2$, and Pt/H adsorption that were not explicitly included in the training set. For fcc Pt, we predict the lattice constant to within $0.1 \%$ of the DFT value and the bulk modulus and elastic constants to within $6\%$ (see Table \ref{bulk_pt}). This considerably improves on a recent ReaxFF Pt/H force field \cite{gai2016atomistic}, which overestimates the $C_{44}$ elastic constant by nearly $200\%$. We next consider a more stringent test of model performance by forcing the model to extrapolate to structures that are significantly different from those encountered during training. In Fig.\ \ref{fig3}, we plot model predictions of bulk Pt energies as a function of volume, hydrogen dissociation and dimer interaction profiles, and surface and binding energies for several Pt surface orientations and adsorption sites. In each case we present the $99\%$ confidence region associated with each prediction, which we compute under the Deterministic Training Conditional approximation of the GP predictive variance (see Methods). In general, we observe low uncertainties and excellent agreement with DFT near configurations that are well-represented in the training set. For bulk Pt, shown in Fig.\ \ref{fig3}a, the uncertainties nearly vanish close to the equilibrium volume, which was extensively sampled in the $0$ GPa training simulation of bulk Pt. The model is found to give confident and accurate predictions for H$_2$ bond lengths between $\sim 0.5$ and $1.2 \text{ \AA}$ (Fig.\ \ref{fig2}b) and for dimer separations above $1.8 \text{ \AA}$ (Fig.\ \ref{fig2}c), with the confidence region growing for extreme bond lengths and dimer separations that were not encountered during training. For the Pt(111) surface energies and binding energies (Figs.\ \ref{fig2}d and e), the model is in excellent agreement with DFT for the (111) surface and extrapolates surprisingly well to other surface orientations. The largest uncertainties are observed for the (110) and (100) hollow site binding energies, likely because there is not a geometrically similar binding site on the Pt (111) surface.

\subsection{Large-scale reactive MD}
\begin{figure*}
	\centering
    \includegraphics{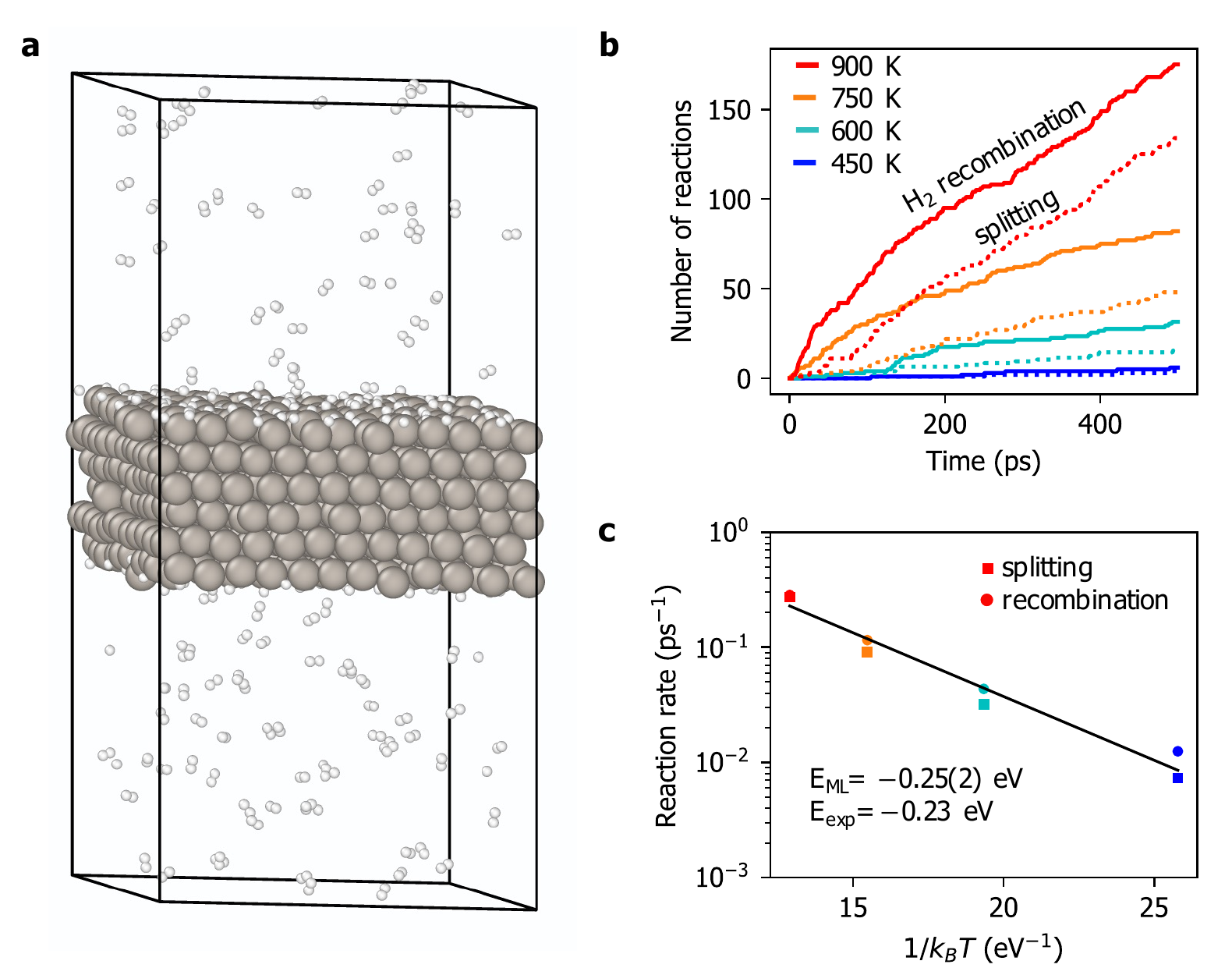}
	\caption{Reactive molecular dynamics with the mapped SGP force field. (a) MD snapshot from a 900 K Pt/H simulation. The system consists of a six-layer 12-by-12 Pt slab in the (111) orientation with H$_2$ molecules splitting and recombining at both Pt surfaces. (b) Cumulative number of reactive recombination and splitting events observed in fixed-temperature MD simulations with the temperature ranging from $450$ to $900$ K. The slope of each curve after an initial equilibration period of $200 \text{ ps}$ is used to estimate the reaction rate of hydrogen splitting and recombination at each temperature. (c) Arrhenius plot of the reaction rate as a function of inverse temperature, yielding an estimated activation energy of $E_{\text{SGP}} = 0.25(2) \text{eV}$ for steady-state hydrogen splitting and recombination on Pt(111).}
	\label{fig4}
\end{figure*}

Finally, we demonstrate that the trained force field can be used to perform large-scale reactive MD simulations that accurately probe surface reactivity. The trained SGP was mapped onto an accelerated quadratic model and used to perform fixed-temperature MD simulations to quantify the reaction rates of hydrogen splitting and recombination on the Pt (111) surface. The initial frame of each simulation consisted of (i) a six-layer 12-by-12 Pt (111) slab with both surfaces fully covered with H adsorbates, and (ii) 80 randomly-oriented H$_2$ molecules in the gas phase, giving 864 Pt atoms and 448 H atoms in total (see Fig.\ \ref{fig4}a for an example MD snapshot). The simulations were performed at four temperatures---450, 600, 750, and 900 K---with the vertical dimension of the box frozen at $68 \text{ \AA}$ and with the pressure along the transverse axes set to $0$ GPa. For each run, $500$ ps of dynamics were simulated in total with a timestep of $0.1$ fs. Reactive events were counted by monitoring the coordination number of each hydrogen atom within a $1$-$\text{\AA}$ shell, which was either zero or one depending on whether the atom was adsorbed on the surface or in the gas phase, respectively. During the first ${\sim 100-200}$ ps of the simulation, the recombination rate exceeded the adsorption rate as the system approached equilibrium surface coverage, with higher temperatures associated with lower equilibrium coverage (see Supplementary Figure 4 for a plot of the surface coverage as a function of time). Once the slab reached equilibrium coverage, the rates of splitting and recombination became roughly equal (see Fig.\ \ref{fig4}b), and the reaction rates were estimated by performing a linear fit of the cumulative reaction count as a function of time for the final $300 \text{ ps}$ of each simulation. The estimated rates from these linear fits are plotted against inverse temperature in Fig.\ \ref{fig4}, giving an estimated activation energy of $0.25(2) \text{ eV}$. To check the effect of pressure, the simulations were repeated with the size of the vacuum doubled, giving a consistent estimate of $0.20(3) \text{ eV}$ for the activation energy. Both estimates are consistent with the experimentally measured value of $0.23 \text{ eV}$ \cite{montano2006hydrogen}.

\section{Discussion}
We have developed a method for training reactive force fields on the fly that achieves excellent accuracy relative to DFT and is competitive with the primary traditional force field used for reactive MD simulations, ReaxFF, in terms of computational efficiency. The method presented here is intended to reduce the time, effort, and expertise required to train accurate reactive force fields, making it possible to obtain a high quality model in a matter of days with minimal human supervision. Our method bridges two of the leading approaches to ML force fields, combining the principled Bayesian uncertainties available in kernel-based approaches like GAP with the computational efficiency of parametric force fields such as SNAP \cite{thompson2015spectral}, qSNAP \cite{wood2018extending}, and MTP \cite{shapeev2016moment}. We hope that by simplifying the training of reactive force fields and reducing the cost of reactive MD, this unified approach will help extend the range of applicability and predictive power of reactive MD as a computational tool, providing new insights into complex systems that have so far remained out of reach of computational modelling. Of particular interest is the application of the method discussed here to biochemical reactions and more complex heterogeneous catalysts.

\section{Methods}
In this section we present the key methods used in the paper. We present our implementation of SGP force fields in \ref{sgps}, our procedure for mapping them onto accelerated polynomial models in \ref{mp}, and the computational details of our MD, ML, and DFT calculations in \ref{details}. Throughout, we use arrows to denote vectors in three-dimensional space, e.g.\ $\vec{r}_1$, and bold font to denote vectors and tensors in more general feature spaces, e.g.\ $\mathbf{d}_i$.

\subsection{Sparse Gaussian process force fields}
\label{sgps}

Our SGP force fields are defined in three key steps:
(i) Fixing the energy model, which involves expressing the total potential energy as a sum over local energies assigned to atom-centered local environments $\rho_i$,
(ii) mapping these local environments onto descriptor vectors $\mathbf{d}_i$ that serve as input to the model, and
(iii) choosing a kernel function $k(\mathbf{d}_i, \mathbf{d}_j)$ that quantifies the similarity between two local environments. In this section, we summarize these steps and then present our approach to making predictions, evaluating Bayesian uncertainties, and optimizing hyperparameters with the SGP.

\subsubsection{Defining the energy model}
As in the Gaussian Approximation Potential formalism \cite{bartok2010gaussian}, we express the total potential energy $E(\vec{r}_1, ..., \vec{r}_N; s_1, ... s_N)$ of a structure of $N$ atoms as a sum over local energies $\varepsilon$ assigned to atom-centered local environments,
\begin{equation}
E(\vec{r}_1, ..., \vec{r}_N; s_1, ... s_N) = \sum_i^N \varepsilon(s_i, \rho_i).
\label{pot_eng}
\end{equation}
Here $\vec{r}_i$ is the position of atom $i$, $s_i$ is its chemical species, and we define the local environment $\rho_i$ of atom $i$ to be the set of chemical species and interatomic distance vectors connecting atom $i$ to neighboring atoms $j\ne i$ within a cutoff sphere of radius $r_{\text{cut}}^{(s_i, s_j)}$,
\begin{equation}
    \rho_i = \{(s_j, \vec{r}_{ij}) | r_{ij} < r_{\text{cut}}^{(s_i, s_j)} \}.
\label{loc_env}
\end{equation}
Note that we allow the cutoff radius $r_{\text{cut}}^{(s_i, s_j)}$ to depend on the central and environment species $s_i$ and $s_j$. This additional flexibility was found to be important in the Pt/H system, where H-H and H-Pt interactions were found to require shorter cutoffs than Pt-Pt.


\subsubsection{Describing local environments}
To train an SGP model, local environments $\rho_i$ must be mapped onto fixed-length descriptor vectors $\mathbf{d}_i$ that respect the physical symmetries of the potential energy surface. We use the multi-element atomic cluster expansion of Drautz \cite{drautz2019atomic, drautz2020atomic, bachmayr2019atomic} to efficiently compute many-body descriptors that satisfy rotational, permutational, translational, and mirror symmetry. To satisfy rotational invariance, the descriptors are computed in a two-step process: the first step is to compute a rotationally \textit{covariant} descriptor $\mathbf{c}_i$ of atom $i$ by looping over the neighbors of $i$ and passing each interatomic distance vector $\vec{r}_{ij}$ through a basis set of radial basis functions multiplied by spherical harmonics, and the second step is to transform $\mathbf{c}_i$ into a rotationally \textit{invariant} descriptor $\mathbf{d}_i$ by tensoring $\mathbf{c}_i$ with itself and keeping only the rotationally invariant elements of the resulting tensor. In the first step, the interatomic distance vectors $\vec{r}_{ij}$ are passed through basis functions $\phi_{n\ell m}$ of the form
\begin{equation}
    \phi_{n\ell m}\left( \vec{r}_{ij} \right) = R_{n}(\tilde{r}_{ij}) Y_{\ell m}(\hat{r}_{ij}) c(r_{ij}, r_{\text{cut}}^{(s_i, s_j)}),
\end{equation}
where $\tilde{r}_{ij} = \frac{r_{ij}}{r_{\text{cut}}^{(s_i, s_j)}}$ is a scaled interatomic distance, $R_{n}$ are radial basis functions defined on the interval $[0, 1]$, $Y_{\ell m}$ are the real spherical harmonics, and $c$ is a cutoff function that smoothly goes to zero as the interatomic distance $r_{ij}$ approaches the cutoff radius $r_{\text{cut}}^{(s_i, s_j)}$. The covariant descriptor $\mathbf{c}_i$ (called the ``atomic base'' in \cite{drautz2019atomic}) is a tensor indexed by species $s$, radial number $n$ and angular numbers $\ell$ and $m$, and is computed by summing the basis functions over all neighboring atoms of a particular species $s$,
\begin{equation}
c_{isnlm} = \sum_{j \in \rho_i} \delta_{s, s_j} \phi_{n\ell m}(\vec{r}_{ij}),
\label{cov}
\end{equation}
where $\delta_{s, s_j} = 1$ if $s = s_j$ and $0$ otherwise. Finally, in the second step, the rotationally invariant vector $\mathbf{d}_i$ is computed by invoking the sum rule of spherical harmonics,
\begin{equation}
d_{i s_1 s_2 n_1 n_2 \ell} = \sum_{m = -\ell}^{\ell} c_{i s_1 n_1 \ell m} c_{i s_2 n_2 \ell m}.
\end{equation}
To eliminate redundancy in the invariant descriptor, we notice that interchanging the $s$ and $n$ indices leaves the descriptor unchanged,
\begin{equation}
d_{i s_1 s_2 n_1 n_2 \ell} = d_{i s_2 s_1 n_2 n_1 \ell}.
\label{inv}
\end{equation}
In practice we keep only the unique values, which can be visualized as the upper- or lower-triangular portion of the matrix formed from invariant contractions of the tensor product $\mathbf{c}_i \otimes \mathbf{c}_i$, as shown schematically in Fig.\ \ref{fig1}(b). This gives a descriptor vector of length $\frac{N_{\text{s}} N_{\text{rad}} (N_{\text{s}} N_{\text{rad}} +1)(\ell_{\text{max}} + 1)}{2}$, where $N_{\text{s}}$ is the number of species, $N_{\text{rad}}$ is the number of radial functions, and $\ell_{\text{max}}$ is the maximum value of $\ell$ in the spherical harmonics expansion. When computing descriptor values, we also compute gradients with respect to the Cartesian coordinates of each neighbor in the local environment, which are needed to evaluate forces and stresses.

The cutoff radii $r_{\text{cut}}^{(s_i, s_j)}$ and radial and angular basis set sizes $N_{\text{rad}}$ and $\ell_{\text{max}}$ are hyperparameters of the model that can be tuned to improve accuracy. We chose the Chebyshev polynomials of the first kind for the radial basis set and a simple quadratic for the cutoff,
\begin{equation}
c(r_{ij}, r_{\text{cut}}^{(s_i, s_j)}) = (r_{\text{cut}}^{(s_i, s_j)} - r_{ij})^2.
\end{equation}
We set the Pt-Pt cutoff to $4.25 \text{ \AA}$ and the H-H and H-Pt cutoffs to $3.0 \text{ \AA}$, and truncated the basis expansion at $N_{\text{rad}} = 8$, $\ell_{\text{max}} = 3$. These hyperparameters performed favorably on a restricted training set of Pt/H structures, as determined by the log marginal likelihood of the sparse GP (see Supplementary Figure 1).

\subsubsection{Making model predictions}
The sparse GP prediction of the local energy $\varepsilon$ assigned to environment $\rho_i$ is evaluated by performing a weighted sum of kernels between $\rho_i$ and a set of representative sparse environments $S$,
\begin{equation}
\varepsilon(\rho_i) = \sum_{s\in S}^{N_s} k(\mathbf{d}_i, \mathbf{d}_s) \alpha_s,
\label{en_pred}
\end{equation}
where $N_s$ is the number of sparse environments, $k$ is a kernel function quantifying the similarity of two local environments, and $\boldsymbol{\alpha}$ is a vector of training coefficients. For the kernel function, we use a normalized dot product kernel raised to an integer power $\xi$, similar to the SOAP kernel \cite{bartok2013representing}:
\begin{equation}
k(\mathbf{d}_1, \mathbf{d}_2) = \sigma^2 \left( \frac{\mathbf{d}_1 \cdot \mathbf{d}_2}{d_1 d_2}\right)^{\xi}.
\label{kern}
\end{equation}
Here $\sigma$ is a hyperparameter quantifying variation in the learned local energy, and in our final trained model is set to $3.84 \text{ eV}$.

The training coefficients $\boldsymbol{\alpha}$ are given by
\begin{equation}
\boldsymbol{\alpha} = \mathbf{\Sigma} \mathbf{K}_{SF} \mathbf{y},
\label{training_eq}
\end{equation}
where $\mathbf{\Sigma} = (\mathbf{K}_{SF} \mathbf{\Lambda}^{-1} \mathbf{K}_{FS} + \mathbf{K}_{SS})^{-1}$, $\mathbf{K}_{SF}$ is the matrix of kernel values between the sparse set $S$ and the training set $F$ (with $\mathbf{K}_{FS} = \mathbf{K}_{SF}^\top$), $\mathbf{K}_{SS}$ is the matrix of kernel values between the sparse set $S$ and itself, $\mathbf{\Lambda}$ is a diagonal matrix of noise values quantifying the expected error associated with each training label, and $\mathbf{y}$ is the vector of training labels consisting of potential energies, forces, and virial stresses. For the noise values in $\mathbf{\Lambda}$, we chose for the final trained model a force noise of $\sigma_F = 0.1 \text{ eV/\AA}$, an energy noise of $\sigma_E = 50 \text{ meV}$, and a stress noise of $\sigma_S = 0.1 \text{ GPa}$. In practice, we found that performing direct matrix inversion to compute $\mathbf{\Sigma}$ in Eq.\ (\ref{training_eq}) was numerically unstable, so we instead compute $\boldsymbol{\alpha}$ with QR decomposition, as proposed in Ref.\ \cite{foster2009stable}.


\subsubsection{Evaluating uncertainties}
To evaluate uncertainties on total potential energies $E$, we compute the GP predictive variance $V(E)$ under the Deterministic Training Conditional (DTC) approximation \cite{quinonero2005unifying},
\begin{equation}
V(E) = k_{EE} - \mathbf{k}_{ES} \mathbf{K}_{SS}^{-1} \mathbf{k}_{ES} + \mathbf{k}_{ES} \mathbf{\Sigma} \mathbf{k}_{SE}.
\label{en_var}
\end{equation}
Here $k_{EE}$ is the GP covariance between $E$ and itself, which is computed as a sum of local energy kernels
\begin{equation}
    k_{EE} = \langle E, E \rangle = \sum_{i, j = 1}^{N} \langle \varepsilon_i, \varepsilon_j \rangle = \sum_{i,j=1}^{N} k(\mathbf{d}_i, \mathbf{d}_j)
\label{en_kern}
\end{equation}
with $i$ and $j$ ranging over all atoms in the structure. The row vector $\mathbf{k}_{ES}$ stores the GP covariances between the potential energy $E$ and the local energies of the sparse environments, with $\mathbf{k}_{SE} = \mathbf{k}_{ES}^\top$. 

Surface energies and binding energies are linear combinations of potential energies, and their uncertainties can be obtained from a straightforward generalization of Eq.\ (\ref{en_var}). Consider a quantity $Q$ of the form $Q = a E_1 + b E_2$, where $a$ and $b$ are scalars and $E_1$ and $E_2$ are potential energies. GP covariances are bilinear, so that for instance
\begin{equation}
    \langle Q, E \rangle = a \langle E_1, E \rangle + b \langle E_2, E \rangle,
\end{equation}
and as a consequence the GP predictive variance assigned to $Q$ is obtained by replacing $k_{EE}$ and $\mathbf{k}_{ES}$ in Eq.\ (\ref{en_var}) with $k_{QQ}$ and $\mathbf{k}_{QS}$, respectively, where
\begin{equation}
    k_{QQ} = \langle Q, Q \rangle = a^2 k_{E_1 E_1} + b^2 k_{E_2 E_2} + 2 a b k_{E_1 E_2}
\end{equation}
and
\begin{equation}
    \mathbf{k}_{QS} = a \mathbf{k}_{E_1 S} + b \mathbf{k}_{E_2 S}.
\end{equation}
We use these expressions to assign confidence regions to the surface and binding energies reported in Fig.\ \ref{fig3}.

To evaluate uncertainties on local energies $\varepsilon$, we first compute a simplified predictive variance
\begin{equation}
V(\varepsilon) = k_{\varepsilon \varepsilon} - \mathbf{k}_{\varepsilon S} \mathbf{K}_{SS}^{-1} \mathbf{k}_{S \varepsilon}.
\label{scaled_var}
\end{equation}
Formally, $V(\varepsilon)$ is the predictive variance of an exact GP trained on the local energies of the sparse environments, and it has two convenient properties: (i) it is independent of the noise hyperparameters $\mathbf{\Lambda}$, and (ii) it is proportional to but otherwise independent of the signal variance $\sigma^2$. This allows us to rescale the variance to obtain a unitless measure of uncertainty $\widetilde{V}(\varepsilon)$,
\begin{equation}
\widetilde{V}(\varepsilon) = \frac{1}{\sigma^2} V(\varepsilon).
\end{equation}
Notice that $\widetilde{V}(\varepsilon)$ lies  between $0$ and $1$ and is independent of the kernel hyperparameters, providing a robust uncertainty measure on local environments that we use to guide our active learning protocol.

\subsubsection{Optimizing hyperparameters}
To optimize the hyperparameters of the SGP, we evaluate the DTC log marginal likelihood
\begin{equation}
    \mathcal{L} = -\frac{1}{2} \log |\mathbf{K}_{SF} \mathbf{K}_{SS}^{-1} \mathbf{K}_{FS} + \mathbf{\Lambda}| - \frac{1}{2} \mathbf{y}^\top (\mathbf{K}_{SF} \mathbf{K}_{SS}^{-1} \mathbf{K}_{FS} + \mathbf{\Lambda})^{-1} \mathbf{y} - \frac{N_{\text{labels}}}{2} \log (2 \pi),
    \label{log_like}
\end{equation}
where $N_{\text{labels}}$ is the total number of lables in the training set. Eq.\ (\ref{log_like}) quantifies the likelihood of the training labels $\mathbf{y}$ given a particular choice of hyperparameters. The first term penalizes model complexity while the second measures the quality of the fit, and hence hyperparameters that maximize $\mathcal{L}$ tend to achieve a favorable balance of complexity and accuracy \cite{bauer2016understanding}. During on-the-fly runs, after each of the first $N_{\text{hyp}}$ updates to the SGP, the kernel hyperparameters $\sigma$, $\sigma_{E}, \sigma_{F},$ and $\sigma_{S}$ are optimized with the L-BFGS algorithm by evaluating the gradient of $\mathcal{L}$. We also use the log marginal likelihood to evaluate different descriptor hyperparameters $N_{\text{rad}}$, $\ell_{\text{max}}$, and $r_{\text{cut}}^{(s_i, s_j)}$ and the discrete kernel hyperparameter $\xi$ (see Supplementary Table 1 for a comparison).

\subsection{Mapping to an equivalent polynomial model}
\label{mp}

Letting $\tilde{\boldsymbol{d}}_i = \frac{\mathbf{d}_i}{d_i}$ denote the normalized descriptor of local environment $\rho_i$, we observe that with the dot product kernel defined in Eq.\ (\ref{en_kern}), local energy prediction can be rewritten as
\begin{equation}
\begin{split}
\varepsilon(\rho_i) &= \sigma^2 \sum_s (\tilde{\boldsymbol{d}_i} \cdot \tilde{\boldsymbol{d}_s})^\xi \alpha_s \\
&= \sigma^2 \sum_{s, m_1, ..., m_{\xi}} \tilde{d}_{im_1} \tilde{d}_{sm_1} \cdot\cdot\cdot \tilde{d}_{i m_\xi} \tilde{d}_{sm_\xi} \alpha_s \\
&= \sigma^2 \sum_{m_1, ..., m_\xi} \tilde{d}_{im_1} \cdot \cdot \cdot \tilde{d}_{im_\xi} \left( \sum_s \tilde{d}_{s m_1} \cdot \cdot \cdot \tilde{d}_{s m_\xi} \alpha_s \right) \\
&= \sum_{m_1, ..., m_\xi} \tilde{d}_{im_1} \cdot\cdot\cdot \tilde{d}_{im_\xi} \beta_{m_1, ..., m_\xi},
\end{split}
\label{mapping}
\end{equation}
where in the final two lines we have gathered all terms involving the sparse set into a symmetric tensor $\boldsymbol{\beta}$ of rank $\xi$. Once $\boldsymbol{\beta}$ is computed, mean predictions of the SGP can be evaluated without performing a loop over sparse points, which can considerably accelerate model predictions for small $\xi$. For $\xi = 1$, corresponding to a simple dot product kernel, mean predictions become linear in the descriptor,
\begin{equation}
\varepsilon_{\xi=1}(\rho_i) = \tilde{\boldsymbol{d}} \cdot \boldsymbol{\beta}.
\label{lin}
\end{equation}
Evaluating the local energy with Eq.\ (\ref{lin}) requires a single dot product rather than a dot product for each sparse environment, accelerating local energy prediction with the SGP by a factor of $N_s$, where $N_s$ is the number of sparse environments. For $\xi = 2$, mean predictions are quadratic in the descriptor and can be evaluated with a vector-matrix-vector product,
\begin{equation}
\varepsilon_{\xi=2}(\rho_i) = \tilde{\boldsymbol{d}}^\top \boldsymbol{\beta} \tilde{\boldsymbol{d}}.
\label{quad}
\end{equation}


\subsection{DFT, ML, and MD details}
\label{details}

DFT calculations were performed with the Vienna Ab Initio Simulation (VASP) package \cite{kresse1994theory, kresse1996efficiency, kresse1996efficient} using projector augmented wave (PAW) pseudopotentials \cite{blochl1994projector} and the Perdew-Burke-Ernzerhof (PBE) exchange-correlation functional \cite{perdew1996generalized}. To calculate ACE descriptors, train SGP models, and map SGPs onto accelerated quadratic models, we have developed the FLARE++ code, available at https://github.com/mir-group/flare\_pp. On-the-fly training simulations were performed with FLARE \cite{vandermause2020fly}, available at https://github.com/mir-group/flare, which is coupled to the MD engines implemented in the Atomic Simulation Environment (ASE) \cite{larsen2017atomic}. Production MD simulations were performed with LAMMPS \cite{plimpton1995fast} using a custom pair-style available in the FLARE++ repository.

\section{Acknowledgements}
We thank Albert P. Bart\'{o}k for helpful discussions and David Clark, Anders Johansson, and Claudio Zeni for contributions to the FLARE++ code. J.V. acknowledges funding support from the National Science Foundation under grants 1808162 and 2003725. Y.X. is supported by the US Department of Energy (DOE) Office of Basic Energy Sciences under Award No.\ DE-SC0020128. C.J.O. is supported by the National Science Foundation Graduate Research Fellowship Program under Grant No. (DGE1745303). J.S.L. and C.J.O. acknowledge support from the Integrated Mesoscale Architectures for Sustainable Catalysis (IMASC), an Energy Frontier Research Center funded by the US Department of Energy (DOE), Office of Science, Office of Basic Energy Sciences (BES) under Award No. DE-SC0012573. J.V., J.S.L. and C.J.O used the Odyssey cluster, FAS Division of Science, Research Computing Group at Harvard University, and the National Energy Research Scientific Computing Center (NERSC) of the U.S. Department of Energy. 
\clearpage

\begin{table}
\begin{center}
\begin{tabularx}{0.8\textwidth}{ 
   >{\raggedright}X 
   >{\centering}X
   >{\centering}X
   >{\centering}X
   >{\centering}X
   >{\centering}X
   >{\centering}X
   >{\centering}X 
   >{\centering\arraybackslash}X }
 \hline
 \hline
 System & $T$ & $\tau_{\text{sim}}$ & $\tau_{\text{wall}}$ &  $N_{\text{atoms}}$ & $N_{\text{struc}}$ & $N_{\text{envs}}$ & $N_{\text{sparse}}$ & $N_{\text{labels}}$ \\ 
 \hline
 H$_2$ & 1500 & 5.0 & 1.2 & 54 & 24 & 1296 & 124 & 4056 \\ 
 Pt(111) & 300 & 4.0 & 1.2 & 54 & 4 & 216 & 87 & 676 \\ 
 Pt & 1500 & 10.0 & 1.7 & 108 & 6 & 648 & 179 & 1986 \\ 
 Pt/H & 1500 & 3.7 & 61.4 & 73 & 216 & 15 768 & 2034 & 48 816\\ 
 Total & - & - & 65.5 & - & 250 & 17 928 & 2424 & 55 534 \\
 \hline
\end{tabularx}
\caption{Summary of the on-the-fly Pt/H training simulations performed in this work. Listed quantities include the temperature $T$ of the simulation (in $\text{K}$), the total simulation time $\tau_{\text{sim}}$ (in $\text{ps}$), the total wall time of the simulation $\tau_{\text{wall}}$ (in hours), the number of atoms in the simulation $N_{\text{atoms}}$, the total number of structures added to the training set $N_{\text{struc}}$, the total number of local environments in all training structures $N_\text{envs}$, the total number of local environments added to the sparse set $N_{\text{sparse}}$, and the total number of training labels $N_{\text{labels}}$.}
\label{training_stats}
\end{center}
\end{table}

\begin{table}[h!]
\begin{center}
\begin{tabularx}{0.8\textwidth}{ 
   >{\raggedright}X 
   >{\centering}X
   >{\centering}X
   >{\centering}X 
   >{\centering\arraybackslash}X }
 \hline
 \hline
 Property & Experiment & DFT & SGP & ReaxFF \\ 
 \hline
 $a$ & 3.923 & 3.968 & 3.970 (0.1) & 3.947 (0.6)\\ 
 $B$ & 280.1 & 248.0 & 258.2 (4.1) & 240.4 (-3.1) \\ 
 $C_{11}$ & 348.7 & 314.5 & 323.2 (2.8) & 332.0 (5.6) \\ 
 $C_{12}$ & 245.8 & 214.8 & 225.7 (5.1) & 194.6 (-9.4) \\ 
 $C_{44}$ & 73.4 & 64.9 & 68.9 (6.2) & 194.6 (199.8) \\ 
 \hline
\end{tabularx}
\caption{Experimentally measured properties of bulk Pt compared with density functional theory, the SGP force field, and ReaxFF. The listed quantities are the lattice constant $a$ (\AA), the bulk modulus $B$ (GPa), and the elastic constants $C_{11}$, $C_{12}$, and $C_{44}$ (GPa). Percent errors relative to DFT are reported in parentheses for the SGP and ReaxFF models.}
\label{bulk_pt}
\end{center}
\end{table}

\clearpage
\bibliography{flare-mb.bib}

\end{document}